# LEVERAGING MACHINE LEARNING TO ENHANCE CLIMATE MODELS: A REVIEW

*Ahmed Elsayed, Shrouk Wally, Islam Alkabbany, Asem Ali, Aly Farag*

CVIP Lab, University of Louisville, USA

## ABSTRACT

Recent achievements in machine learning (ML) have had a significant impact on various fields, including climate science. Climate modeling is very important and plays a crucial role in shaping the decisions of governments and individuals in mitigating the impact of climate change. Climate change poses a serious threat to humanity; however, current climate models are limited by computational costs, uncertainties, and biases, affecting their prediction accuracy. The vast amount of climate data generated by satellites, radars, and Erath System Models (ESMs) poses a significant challenge. ML techniques can be effectively employed to analyze this data and extract valuable insights that aid in our understanding of the Earth's climate. This review paper focuses on how ML has been utilized in the last 5 years to boost the current state-of-the-art climate models. We invite the ML community to join in the global effort to accurately model the Earth's climate by collaborating with other fields to leverage ML as a powerful tool in this endeavor.

*Index Terms*— Machine Learning, Climate Change, Climate Models,

## 1. INTRODUCTION

The study of the earth's climate has gotten man's attention since the ancient ages. However, the first scientific step toward climate modeling was taken by Svante Arrhenius, in 1898, when laid out the theory of the greenhouse effect and concluded that doubling the atmospheric $CO_2$ would raise the Earth's temperature by 5°-6° C [3, 4, 6]. In 1916, Vilhelm Bjerknes introduced the set of motion equations of the atmosphere using the theory of fluids [5, 7]. Later in 1922, Lewis F. Richardson proposed the first numerical technique for systematic forecasting [10] based on Bjerknes' work. this method suffered from numerical instabilities and took roughly three months to forecast weather during a single day. He also proposed a method for parallelizing calculations to increase throughput. The first successful Numerical Weather Prediction (NWP) experiment came about in 1950 when Von Neumen recruits Jule G. Charney to develop a numerical framework for weather prediction [11]. In 1956, Norman Phillips developed the first Atmospheric General Circulation Model (AGCM) [12]. From 1955 to 1565, the first 3D atmosphere model was developed, leading to the Geophysical Fluid Dynamical Lab (GFDL) family of GCMs [13]. At the same time, the UCLA family of GCMs was developed[14]. From here there was an explosion in the research related to GCM which led to a complex tree of AGCMs (refer to [15] for more details). These models were originally developed to understand the general circulation of the atmosphere. Later ocean and cryosphere models were developed. To date, there are two predominant models used for climate predictions, General Circulation Models (GCMs) and Coupled Earth System Models (CESMs) [16]. These models are used as predictive models on timescales of days to weeks (weather predictions) up to centuries or millennia (long-term climate projections) and play a crucial role in shaping the decisions made by both local and national governments, as evidenced by the Intergovernmental Panel on Climate Change (IPCC) reports [4,5]. Additionally, these models assist individuals in assessing their climate-related risks.

Climate Models (CMs) can run on modern computer systems through discretization by dividing the planet into many small regions called grid cells. These cells can capture processes within each region and processes that enable different regions to communicate. However, current CMs are computationally expensive even when running on the most advanced supercomputers [18] (simulations may take weeks to run). Additionally, these models suffer from uncertainties, and biases, which affect the accuracy of their predictions. Tons of climate data are available either via satellites [19, 20], ground stations, or climate modeling projects [21] (simulated climate data). ML models can be trained fast and, they are less computationally extensive. There are many opportunities to use data from diverse fields and remote sensing sources and apply ML algorithms on these data to get important information, see [22-24]. In general, ML models can be more accurate or less expensive than other physical models in two scenarios: (1) the system physics is ambiguous, hard to model and there is plenty of collected data that describes the system. (2) there is an accurate model, but it is too computationally expensive.

The rest of this paper is dedicated to a review of the scientific papers published in the last 5 years addressing the utilization of ML methods and algorithms in climate modeling and analysis. This topic is large and diverse, so the paper is divided into sections to make it easy for the reader to get the most out of it (see figure 1).

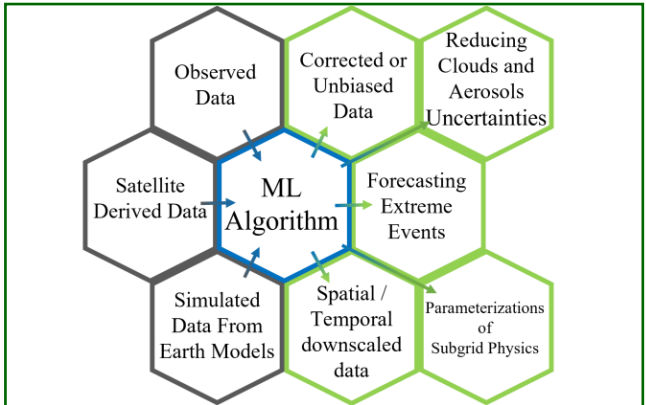

**Fig. 1 A summary of the interaction between various subdomains of climate modeling and ML**.

Section 2 intends to establish a connection between data, ML, and climate modeling by bridging the gap that currently exists between them. This section has three subsections: section 2.1 is dedicated to the parameterization of subgrid atmospheric processes and the physics-informed ML for climate modeling, section 2.2 focuses on reducing uncertainties associated with clouds and Aerosol with the help of ML, and section 2.3 is about downscaling of outputs of climate models using ML techniques. Section 3 aims at detecting and forecasting extreme events. Section 4 delves into the current and future roles of ML in climate modeling and presents a discussion and conclusion on the subject. Lastly, we show a straightforward sample work done at the CVIP lab related to climate modeling.

## 2. DATA AND ML CLIMATE MODELING

The Earth's climate can be studied through data collection, which can then be transformed into practical predictions by condensing it into coherent and computationally tractable models. ML models are often more precise or cost-effective compared to other models in cases where there is ample data that is difficult to model using traditional statistics, or when existing models are too computationally intensive for practical use in production.

**2.1 Parameterizations and physics informed ML for CM**

CMs harvest our knowledge of Earth and climate physics. Roughly they can be divided into two components, the Dynamical Core, and the Physical Parameterizations. The Dynamical Core is responsible for solving a system of coupled partial differential equations that govern the atmospheric flow and so handles transport and exchange between cells. the Physical Parameterizations handle all the processes not captured by the flow equations including moisture, ice, etc. These models are accurate but computationally expensive. ML can provide techniques for solving such systems efficiently and this can accelerate the current CMs [25].

The complexity of ocean mixing processes poses a significant challenge, and traditional physics-based parameterizations have not yielded satisfactory results in accurately representing them [26]. As a result, these parameterization methods have led to biases in both ocean and climate modeling [27]. There is a pressing need to advance our understanding of ocean mixing and to develop more effective parameterization techniques that can better capture the complexity of these processes and minimize modeling biases. DL has emerged as a valuable tool for accurately parameterizing sub-grid processes, employing a data-driven approach that incorporates the underlying physical principles. An interesting application of this approach involved training a Neural Network (NN) using a decade-long time series of hydrographic and turbulence observations in the tropical Pacific to develop parameterizations of oceanic vertical-mixing processes [28]. The results of this study were encouraging, demonstrating the feasibility of constructing a physics-informed deep-learning parameterization using limited observations and well-established physical constraints. This approach has great potential to enhance the accuracy of climate simulations and improve our understanding of ocean mixing processes.

Recent developments in differential equation-inspired DL techniques attracting considerable attention from the research community. Recent research has demonstrated the equivalence between residual networks and the Euler method for solving ordinary differential equations (ODEs) [29]. Building upon this work, a group of researchers proposed a novel CM that utilizes neural ODEs and the diffusion equation, which they named the neural diffusion equation (NDE) [30]. This innovative approach starts a brand-new application area in climate modeling. Other research employs relevance vector machines (RVMs) and CNNs, to derive computationally efficient ocean mesoscale eddies and parameterizations from data, which are interpretable and/or encapsulate physics [31].

**2.2 Clouds and Aerosol Uncertainties**

Aerosols have the potential to alter cloud properties and precipitation, leading to indirect impacts on the Earth's radiation budget and contributing to climate change. The radiative forcing associated with aerosol-cloud interactions (RFACI) is a primary source of uncertainty in climate prediction. Despite their potentially significant impact on climate, the understanding of aerosol-cloud interactions (ACIs) is limited, resulting in uncertainty in climate projections. To enable computationally efficient numerical weather and CMs, cloud structure is often simplified, which can affect the accuracy of predictions. Hence, reducing uncertainty in RFACI is crucial to improve the accuracy of climate predictions and a better understanding of the role of ACIs in regulating cloud properties and influencing the Earth's radiation budget.

Recent research has demonstrated the potential of using ML models to solve these problems effectively [32, 33]. For instance, A study utilizing random forests has discovered that the impact of aerosols on climate forcing is primarily due to an increase in cloud cover [34]. The study's findings contrast with previous beliefs that aerosols brighten clouds by reducing droplet size. The results provide significant observational evidence that will aid in improving climate models and better representing the impacts of ACIs.

Another study aimed to reduce uncertainties in CMs by improving the calculation of particle number concentration (PNC), an important parameter affecting RFACI, using ML tools [35]. The researchers developed a Random Forest Regression Model that learns aerosol microphysics. When this model has been incorporated into the GISS-ModelE2.1[36], it resulted in a decrease in RFACI from -1.46 to -1.11 W/m2.

Meyer et al. [37] proposed to correct the European Centre for Medium-Range Weather Forecasts 1D radiation scheme for 3D cloud effects using computationally cheap NNs. The networks used the output of the 1D Tripleclouds [38] solver as an input while using the output of the 3D SPARTACUS [39] solver as a target. The networks were able to improve the accuracy of the 1D solver for the 3D cloud effects, with typical errors of 20-30% of the 3D signal, while only increasing runtime by about 1%. Replacing certain CM components with NNs approximators may therefore improve both the cost and the accuracy of global CMs. However, further research is needed to identify other impactful components of CMs that could be replaced by NNs, optimize these models, and automate their training workflows [40].

## 2.3 Downscaling and Bias Correction for Climate Models

Providing accurate climate projections at high spatial and temporal resolutions poses a major and persistent challenge for climate science, as such projections are crucial for informing localized adaptation measures and preparedness for extreme events in the future. Unfortunately, these projections are hindered by computational limitations that prevent us from obtaining the necessary high-resolution data[41]. While present-day GCMs can be economically run for extended periods with spatial resolutions of 50 km or more, they rely on physical parameterizations for subgrid-scale processes such as cumulus convection and gravity wave drag, which are major sources of uncertainty [42]. This uncertainty leads to different regional patterns of climate change being projected when the same model is run at a finer resolution. Moreover, coarse-grid simulations suffer from spatial resolution trade-offs that often fail to accurately represent critical processes like rainfall compared to finer grid runs [43]. Despite ongoing investment and development in GCMs, their typical resolution limits their applicability to local scales. Therefore, Regional CMs (RCMs) were developed to bridge this resolution divide [44]. However, running RCMs at high spatial resolution can be computationally expensive, leading to a limited number of GCM/RCM pairs being used for regional projections, which in turn increases the risk of the underrepresenting initial condition and structural model-based uncertainties in RCM-based projections [45].

To address the need for computationally efficient downscaling methods to supplement projections based on RCMs and GCMs, traditional statistical downscaling and bias correction techniques have been developed [46]. Concurrently, geoscientists have been increasingly drawn to the promising field of ML, particularly deep learning (DL), to develop more advanced downscaling techniques. Recent studies have demonstrated promising advancements in downscaling temperature and precipitation over Europe through the application of Convolutional Neural Networks (CNNs) [47, 48]. Furthermore, other researchers have utilized interpretable DL techniques to achieve High-resolution downscaling of rainfall extremes in New Zealand [49]. In addition, NNs and random forests have been successfully trained to correct a 200 km Resolution CM in multiple climates using 25km resolution simulations [50]. These ML models incorporate state-dependent corrections to the temperature and specific humidity tendencies and predict surface radiative fluxes to accurately adjust single timestep tendencies of the coarse model to align with those of a fine-grid reference simulation. An additional study introduced a novel CNN architecture, known as ConvMOS, which takes into account the systematic and location-specific errors that are commonly observed in CM precipitation estimates [51]. By applying the ConvMOS models to the output of RCM REMO, researchers were able to successfully reduce errors in simulated precipitation. The results demonstrated that a combination of per-location model parameters, which focus on minimizing location-specific errors, and global model parameters, which target systematic errors, can effectively improve model output statistics (MOS) performance. Overall, this approach has shown great potential for enhancing the accuracy of precipitation estimates in CMs.

## 3. MID-RANGE EXTREME EVENT FORECASTING

Our lives are influenced by climate conditions in numerous ways, and therefore, it is crucial to comprehend and forecast extreme weather events associated with climate. Droughts, flash floods, storms, and other natural hazards can have severe consequences on our health, economy, and environment. Hence, by understanding and forecasting these events, we can improve hazard management and minimize the risks they pose. It is crucial to recognize that weather forecasting and climate forecasting are distinct methods with different objectives. Weather models predict short-term atmospheric conditions, typically ranging from a few days to a couple of weeks. CMs, on the other hand, focus on long-term changes in the Earth's climate, ranging from several months to several decades or even centuries [52]. By understanding these differences, we can make more informed decisions and plans for both the short and long term, considering the varying time scales involved in weather and climate forecasting. ML is an exciting technology that holds great promise for reducing the time scale gap between weather and climate forecasts. By analyzing vast amounts of data and identifying complex patterns in atmospheric conditions, ML algorithms can produce reliable forecasts over weeks to a few months [53, 54]. This section demonstrates how ML can help the current Numerical Weather Prediction (NWP) and CM produce accurate extreme weather predictions.

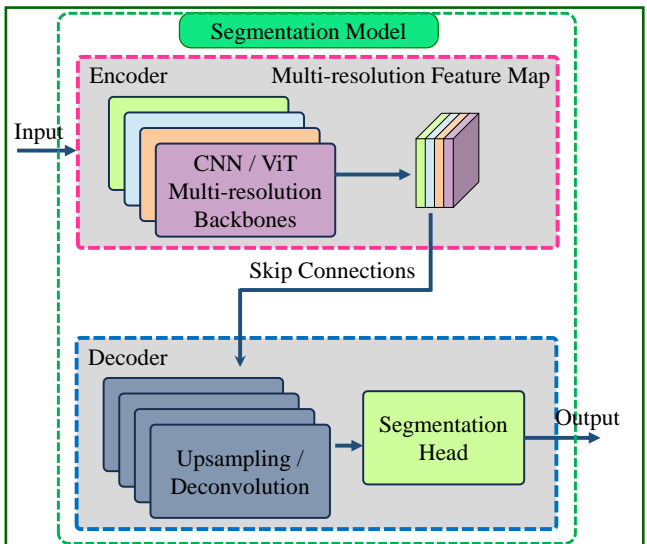

**Fig. 2**: A generic architecture for contrails segmentation; it consists of encoder and decoder. The encoder has backbones that can extract multiscale feature maps. The decoder has multiple upsampling/deconvolution blocks alongside the segmentation head. The details of the encoder and decoder and the skip connection depend on the segmentation model itself (see, [1-5] ).
We used two different backbones for each model either a CNN based or a transformer based.

### 3.1 Local forecasts

All CMs have a coarse grid resolution (~50km), which means that their prediction represents the averages about a certain region. However, for these predictions to be useful they must be specific and tailored to a particular location. ML has been utilized to produce localized forecasts. A recent study used NN to predict the drought occurrence in Munich and Lisbon, with a lead time of one month [55]. The approach considers 28 atmospheric and soil variables as input parameters from the CRCM5-LE model [56], this method correctly classified drought or no drought for around 55%-57% of the events for both domains. Another study used NN to predict floods at times of extreme storms [57]. In a study conducted by a separate team of researchers, various ML algorithms namely, Random Forests, XGBoost, NN, and LSTM were trained on the output of the CESM-LENS model [58] to predict seasonal precipitation patterns across the western United States [59]. The results of their analysis demonstrated that these ML techniques can compete with, and in some cases even surpass, the forecasting capabilities of existing dynamical models in the North American Multi-Model Ensemble.

### 3.2 Storm detection and tracking

Identifying extreme events in CM output is a classification challenge because the available datasets are strongly unbalanced. This is true because extreme events by definition are rare. ML has been successfully utilized to classify extreme weather events such as cyclones, atmospheric rivers, and tornadoes in historical climate datasets using deep learning techniques [60-63]. It is crucial to develop supplementary tools for a broader range of event categories, online tools that integrate with CMs, annotated datasets for predicting forthcoming events, and statistical tools to measure the uncertainty in new extreme event predictions.

## 4. EXPERIMENT

As an example, for the ML research at the CVIP Lab related to CM, we utilized deep learning models for contrails detection and segmentation from satellite imagery. Contrails, or vapor trails, are elongated and white lines that form in the wake of an aircraft's engines when they interact with the extremely cold and moist upper atmosphere. These trails consist primarily of ice crystals, water droplets, and particulate matter emitted from the aircraft's exhaust.

Determining the overall impact of contrails on global warming remains an ongoing research challenge. Studies indicate that contrails significantly contribute to the overall climate impact of aviati. Contrails contribute as much to global warming as the fuel they burn for flight, approximately 1% of all human-caused global warming.

Learning, advancements in satellite technology and deep learning algorithms can revolutionize the study of contrails by accurately identifying and tracking them amidst the vast amounts of satellite data available. This will give researchers valuable insights into contrail formation, their persistence, and their potential implications for global warming.

### 4.1 Problem description:

Contrails can be detected in satellite imagery encompassing different bands capturing various wavelengths of light. Thermal infrared bands are particularly valuable for detecting contrails during both day and night, as they highlight the temperature difference between contrails and their surroundings. Satellites can only capture contrails that have a width larger than 1 km since most satellites have a resolution of 1 pixel/km. so, our goal is to use a deep learning model to accurately identify and segment aviation contrails in geostationary satellite images using infrared (IR) bands. This will provide valuable insights into the impact of contrails on climate change.

### 4.2 Proposed approach

#### 4.2.1 Dataset:

For this experiment, we will be using a dataset of satellite images obtained from the GOES-16 Advanced Baseline Imager (bands 8-16). The dataset details can be obtained from [64].

#### 4.2.2 Data Preparation

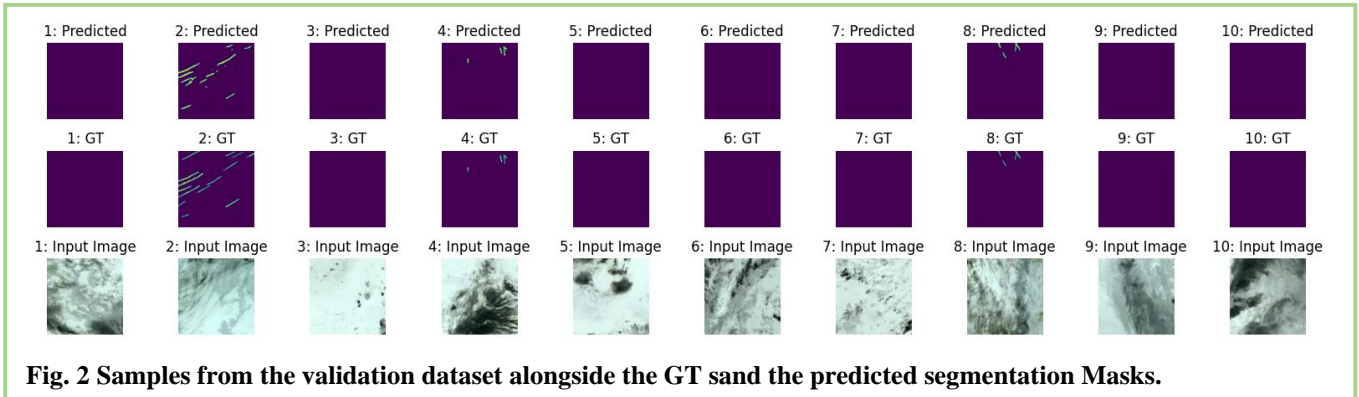
**Fig. 2 Samples from the validation dataset alongside the GT sand the predicted segmentation Masks.**

The dataset contains multispectral images with 9 different bands, not all of them are sensitive to contrails formation. According to the NOAA ABI satellite imagery and products guide, bands 11, 13, 14, and 15 are most influenced by contrails. Therefore, we will utilize only these four bands to create a false color image that maximizes the appearance of contrails. The false color image will have 5 channels; the first three channels are the same as the Ash color scheme[65] while the last two channels are the last two channels from the dust color scheme [65]. Combining these two color schemes, we aim to enhance the visibility and distinguishability of contrails in satellite imagery, facilitating more accurate and efficient detection of contrails in our model.

### 4.2.3 Model Architecture

Figure 2 shows a generic architecture for contrails segmentation. This architecture has an encoder and a decoder. The encoder has multiple blocks known as backbones that are used to extract multi-resolution features from the input image. These feature maps are then passed to the upsampling/Deconvolution blocks of the decoder via skip connections. The resulting feature maps are then passed to a segmentation head to generate the final segmentation mask. Using such architecture enables us to experiment with multiple segmentation models using different backbones. Namely, we experimented with 5 different segmentation models: Unet++ [1], Linknet [2], PAN [3], MAnet [4] and DeepLabV3+ [5]. There are many backbones we can choose from and generally, we can split them into two categories, CNN-based and Transformer-based. To limit the number of experiments we picked one backbone from each category that has the largest of learning parameters. This choice makes sense because as the number of parameters increases the model's capacity to learn complex tasks increases. This leaves us with 1o different models (5 models x 2 backbones, see Table 1 for the exact names of the backbones). We used models and the encoders available from the segmentation Models Pytorch library [66].

Training the models:

All 10 models were trained for 100 epochs using the AdamW optimizer [67], with the Cosine Annealing learning rate scheduler [68]. Only the training dataset was used for training while the validation dataset was used for validation only. The Dice Loss was utilized as a loss function and the Dice Score was utilized as a validation metric to evaluate the performance of the model [64]. We saved only the model with the best performance based on the dice score and used that model for testing.

Results:

Table 1 shows the dice score for the 10 models we experimented with. You can see that Unet++ [1] with tu-resnest269e backbone gives the best results despite being quite old. Figure 3 shows samples from the validation dataset with the ground truth segmentation mask and the predicted one. You can see that the network was able to learn to detect and segment contrails from satellite images. Please note that in Figure 3, we used the Ash color scheme for only visualizing the input since the original input image has 5 channels and we cannot visualize it.

## 5. FUTURE DIRECTIONS

Climate Modeling is a complex problem that requires accurate CMs to inform policies and decisions. ML techniques offer a promising approach to tackling the challenges associated with climate modeling. In this article, we reviewed some of the recent advancements in ML for climate modeling, including parameterizations and physics-informed ML for CM, clouds, and aerosol uncertainties, and downscaling and bias correction for CMs. The reviewed

**Table 1 comparison of different models' performance on the validation dataset.**

| Segmentation Model | Backbone | Dice Score |
|---|---|---|
| Unet++ [1] | tu-resnest269e [8] | 65.63 |
| | maxvit_xlarge_tf_512 [9] | |
| MAnet [4] | tu-resnest269e [8] | |
| | maxvit_xlarge_tf_512 [9] | |
| Linknet [2] | tu-resnest269e [8] | |
| | maxvit_xlarge_tf_512 [9] | 64.33 |
| PAN [3] | tu-resnest269e [8] | |
| | mobilevitv2_200 [17] | |
| DeepLabV3+ [5] | tu-resnest269e [8] | 60.38 |
| | mobilevitv2_200 [17] | |

literature demonstrates that ML models can be more precise or cost-effective compared to traditional models in cases where there is ample data that is difficult to model using traditional statistics, or when existing models are too computationally intensive for practical use in production.

However, more work is needed to address the challenges of identifying extreme events in CM output, developing additional tools for more event types, and quantifying the uncertainty in new extreme event forecasts. The future of extreme weather forecasting may involve a collaborative effort between human experts and automated forecasts to improve the accuracy of predictions and better prepare for extreme weather events.